# Reconfigurable 3D magnonic crystal:
# tunable and localized spin-wave excitations in CoFeB meander-shaped film

A.V. Sadovnikov,[1*] G. Talmelli,[2,3*] G. Gubbiotti,[4#] E.N. Beginin,[1] S. Sheshukova,[1] S. A. Nikitov,[5] C. Adelmann,[2] and F. Ciubotaru[2]

[1] Laboratory Metamaterials, Saratov State University, 83 Astrakhanskaya street, Saratov 410012, Russia

[2] Imec, 3001 Leuven, Belgium

[3] Departement Materiaalkunde, KU Leuven, 3001 Leuven, Belgium

[4] Istituto Officina dei Materiali del CNR (CNR-IOM), c/o Dipartimento di Fisica e Geologia, Università di Perugia, I-06123 Perugia, Italy.

[5] Kotel'nikov Institute of Radioengineering and Electronics, Russian Academy of Sciences, Moscow 125009, Russia

**Abstract**

In this work, we study experimentally by broadband ferromagnetic resonance measurements, the dependence of the spin-wave excitation spectra on the magnetic applied field in CoFeB meander-shaped films. Two different orientations of the external magnetic field were explored, namely parallel or perpendicular to the lattice cores. The interpretation of the field dependence of the frequency and spatial profiles of major spin-wave modes were obtained by micromagnetic simulations. We show that the vertical segments lead to the easy-axis type of magnetic anisotropy and support the in-phase and out-of-phase spin-wave precession amplitude in the vertical segments. The latter could potentially be used for the design of tunable metasurfaces or in magnetic memories based on meandering 3D magnetic films.

# Corresponding author: Gianluca Gubbiotti
Email: gubbiotti@iom.cnr.it

**Introduction**

During the last few decades, current semiconductor information technologies have faced the problem of the next miniaturization steps along with the limitations of Joule heating and increasing frequency range.[1,2,3] The progress in the fabrication of magnetic periodic micro and nanostructures, the so-called magnonic crystals (MCs),[4,5,6] improve the possibility for tuning the frequencies of the magnetization oscillation and rich spectra of spin-wave (SW) phenomena.

To integrate magnonic nanodevices in the magnonic networks requires the easy fabrication of interconnections between the functional blocks localized in the different layers of the whole magnonic circuit. [3,7] The concept of 3D magnonics looks promising in this respect, as it offers numerous solutions for the realization of vertical interconnections. Recent studies on 3D magnonic structures based on metallic (CoFeB, NiFe) [8,9] and dielectric (YIG)[10,11] materials have shown that it is possible to bend SWs at the film segments located at angle of 90° to each other and transmitting the SW signal from different vertical heights. These studies report the relatively strong influence of the profile of the substrate on the resulting SW properties. Brillouin light scattering experiments (BLS) on CoFeB and CoFeB/Ta/NiFe meandering films with a period in the submicron range show both propagating and stationary modes as well as frequency band gaps. [8,9] Besides the absence of the band gap at the edge of the Brillouin zones ($k_{BZ}=n\pi/a$) with odd index *n,* which is related to the gliding plane symmetry of meander structure, [8,9] the excitation of SW modes with defined dynamic magnetization profiles is also an important issue. Meander-shaped YIG structure with micrometer-sized period manifested its 3D nature in the strong spin-wave excitation peaks observed by ferromagnetic resonance (FMR) technique. At the same time, the applied bias magnetic field orientation could bring additional properties for potential applications of the meander structures, since the variation of the magnetic field orientation is associated with the reconfiguration of the ground state of the confined magnetic structure, leading to a transformation of the SW excitation spectra.

Compared to isolated or coupled magnetic stripe arrays, [12,13] the meander-shaped structure may exhibit different static and dynamic magnetization behavior due to the vertical segments connecting the horizontal ones. While the localized modes arise from SW excitations localized at the edges of the stripe, [14] central modes arise from the quasi-uniform precession of the magnetization in the central part of the stripe, which explains the different frequency ranges. The vertical segments of the meandering film are more attractive for controlling the localized SWs modes, since not only the value of the magnetic field but also its orientation will affect the SW behavior. Moreover, the SW spectra at zero wavenumbers could allow the 3D magnonic crystals to be used as a metasurface for electromagnetic waves incident on the top of the magnetic meander-shaped film. The formation of a reconfigurable metasurface exploiting the value and orientation of the external magnetic field could help to fabricate absorbing or reflecting layers for electromagnetic waves in the GHz frequency range.

In this work, we show that 3D MCs consisting of a meander-shaped ferromagnetic CoFeB film introduces new degrees of freedom to control the SW excitation spectra by the design of the vertical segments which connect horizontal segments placed at different heights. Two orientations of the applied magnetic field were considered, namely parallel and perpendicular to the groove edges, since the variation of the bias field angle in meander structure can be used for tunability of the resonance characteristics for such types of metasurfaces.

**Sample fabrication**

The CoFeB meander-shaped film was fabricated on a Si wafer using 300 mm industrial platform. In the first step, a 250 nm thick $SiO_2$ layer is formed by controlled thermal oxidation of the Si substrate. Then, the $SiO_2$ is patterned by 248 nm-DUV lithography and conventional reactive ion etching chemistry to form a periodic meander-like surface. The resulting periodic lattice has a height of 50 nm and a width of 300 nm, and a lattice constant of $a=600$ nm. Ta(2nm)/$Co_{40}Fe_{40}B_{20}$/Ta(2nm) films were then deposited onto the lattice by physical vapor deposition. The two Ta layers serve as seed and protection to prevent the oxidation of the ferromagnetic CoFeB film. The typical conformality of the physical vapor deposition process resulted in the film thickness on the vertical segments being about half that on the horizontal ones. [8,9,15]. The saturation magnetization of CoFeB was estimated to be $M_s=1.275$ MA/m using the vibrating sample magnetometry (VSM).

Fig. 1 (a) shows the schematic of a unit cell of the meander-shaped CoFeB film with periodicity $a=L_1+2L_2=600$ nm, where $L_1$ and $L_2$ are the lengths of the horizontal segments with thickness $d_1=23$ nm. The vertical segments have width $d_2=12$ nm and depth of $s=76$ nm. Scanning electron micrographs (SEM) of meander-shaped CoFeB film are presented in Fig. 1 (b) and (c) showing the conformity described above. To note that the horizontal and vertical segments are almost orthogonal and have thicknesses of $d_1=23$ nm and $d_2=12$ nm, respectively.

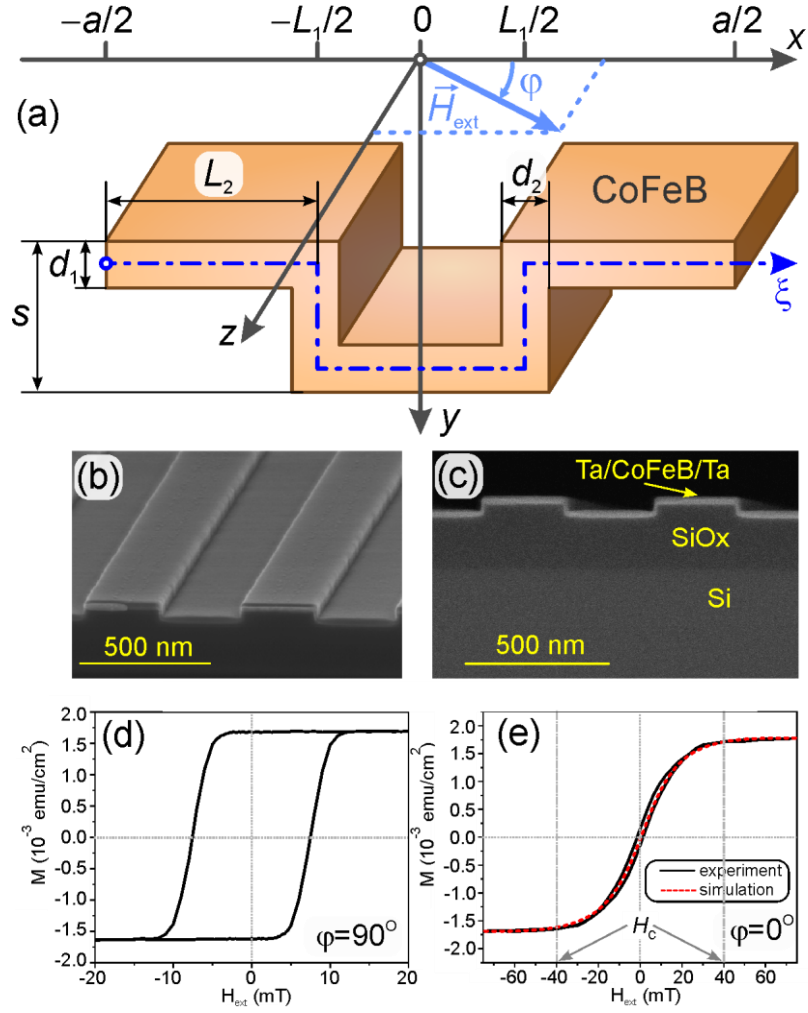

**Fig.1.** (a) Schematic drawing of the meander-shaped film unit cell used in the micromagnetic simulations. (b,c) SEM images of the meander-shaped Ta/CoFeB/Ta film. Panels (d) and (e) show the measured VSM loops for the case where the external magnetic field is aligned along and across to MC lattice cores, respectively.

The dash-dot line represents the coordinate $\xi$ along with the meander structure. An external magnetic field is applied in the $xz$-plane at an angle $\varphi$ to the $x$-axis. The edge of the Brillouin zone (BZ) in reciprocal space corresponds to $\pi/a=0.52\times10^7$ rad/m of this meander-shaped structure. [8,9] In the present study we focus on the case of zero wavenumber (k=0) for SWs in meander-shaped CoFeB film.

**Experiment**

The magnetization loops were measured using the VSM technique at room temperature. Fig. 1 (d) and Fig. 1 (e) show VSM loops for the case where the external field is directed parallel ($\varphi=90°$) and perpendicular ($\varphi=0°$) to the groove cores, respectively. For $\varphi=90°$, the hysteresis loop is almost squared with a coercive field of $H_c \sim 7$ mT and a saturation field of about 12 mT, while for φ=0° almost no coercivity was measured. In the latter case, the loop is linearly dependent on the applied field and reaches saturation at an applied field of about 40 mT.

FMR measurements were performed at room temperature using a vector network analyzer (VNA) and a coplanar waveguide (CPW). The sample is placed between the poles of an electromagnet which provides a magnetic field parallel to the sample surface. [13] The CPW had an inner conductor width of 50 µm and a length of 6.6 mm. The output power was 0 dBm. The sample was placed flip-chip on top of the 50- matched CPW.[16] The RF magnetic field generated by the CPW is responsible for the excitation of the ferromagnetic resonance in the magnetic system under investigation. The $S_{11}$ parameter of the VNA was measured as a function of the frequency $f$ varying between 10 MHz and 40 GHz, and of the magnitude of the applied magnetic field $H_{ext}$. We first saturated the sample along the CPW and thereafter reduced the magnetic field down to zero. Fig. 2 displays the results of experimental measurement (left panels) and numerical simulation (right panels). Fig. 2 (a) and Fig. 2(c) show the results of the VNA-FMR experiment as a color-coded map ($H_{ext}, f$) of $d|S_{11}|/dH_{ext}$ for the case where $H_{ext}$ is directed along the cores (φ=90°) and across them (φ=0°), respectively. The multi-mode SW excitation spectra consist of monotonically increasing $f(H_{ext})$ branches for the case of φ=90° and non-monotonic curves with pronounced minima for φ=0°. For the second case, as the magnetic field decreases from high values saturation, a decrease in frequency is observed, followed by a clear minimum at a certain critical field ($H_{crit}$) and finally by a sudden increase in the mode frequencies $H<H_{crit}$, i.e. in the unsaturated region of magnetization.

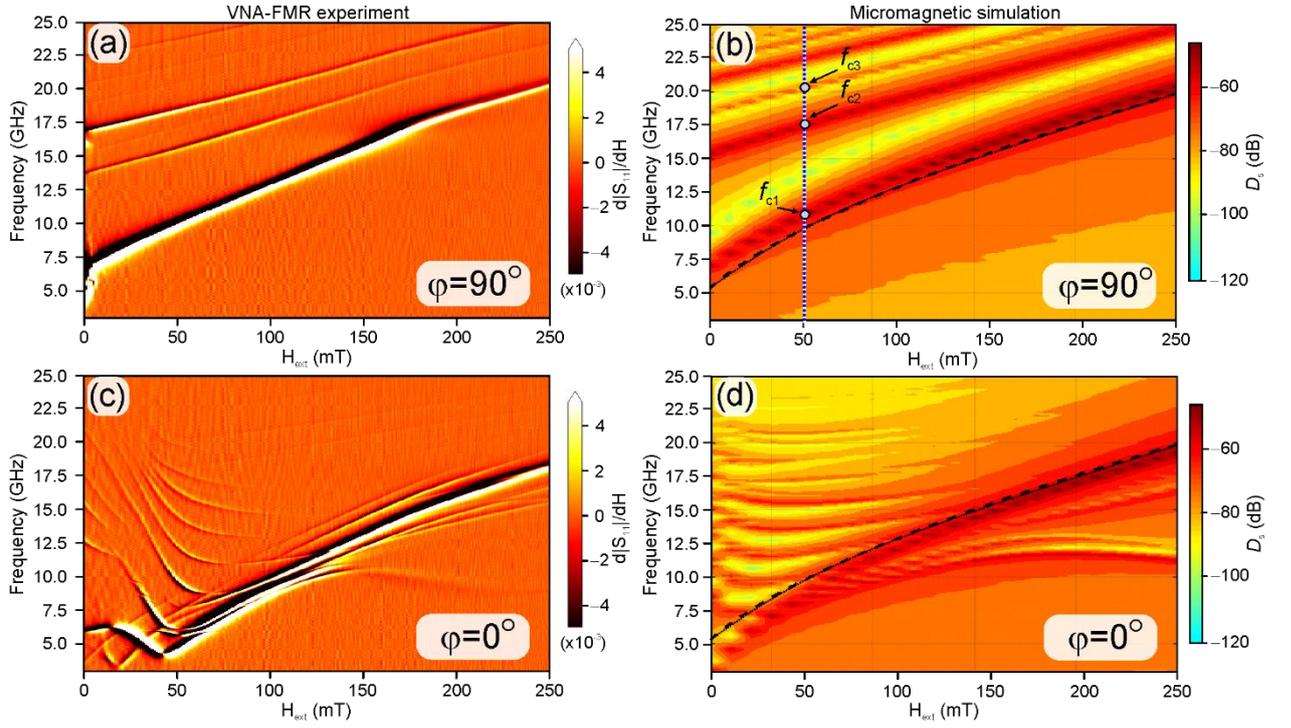

**Fig. 2.** Experimental data from VNA-FMR setup (a,c) and micromagnetic simulation (b,d) for the meander-shaped structure. The external magnetic field is directed either along with the cores ($\varphi=90^O$) in panels (a,b) or transversal to them ($\varphi=0^O$) in panels (c,d).

**Numerical simulation and discussion**

To get a detailed comprehension of the magnetic properties of the meander-shaped CoFeB film, we start calculating the static magnetization distribution using the micromagnetic simulations MUMAX3 micromagnetic software.[17] The simulated structure consists of a period of the meander structure (primitive cell) as shown in Fig. 1. In addition, periodic boundary conditions (PBC) along the $x$- and $z$-axes were considered. Along the $y$-axis, the size of the computational domain corresponded to the height $s$ of meander structures. The simulations were carried out using a cell size of $\Delta x \times \Delta y \times \Delta z = 1 \times 1 \times 1 nm^3$. The simulation of the hysteresis loop for $\varphi=0°$ agrees well with the experimental data, as shown in Fig. 1(e). To simulate the dynamic properties, i.e. the frequency and the spatial profiles of different SW excitations, we varied the external magnetic field $\mu_0 H_{ext}$, aligned along either the $z$- ($\varphi=90^O$) or the $x$-axis ($\varphi=0^O$), from 250 mT to 10 mT in steps of 10 mT, and determined the static magnetization configuration for each step. We have used spatially uniform magnetic field pulses directed along the y-axis of the type of sinc-function with an amplitude of 1 mT and a cut-off frequency of 27 GHz to generate magnetization oscillations in the system.

The spectral power densities $D_s(f)$ of the excited SWs in the meander the unit cell were calculated for each value of the external magnetic field by the Fourier transforming in time of the $y$-component of the magnetization averaged over the unit cell volume $\hat{m}_y(t)$ as $D_s(H_{ext}, f) = 20\log_{10}|\Phi(\hat{m}_y(t))|$, where $\Phi$ is the Fourier transform operator. A time window of 500 ns was considered. The method of calculating the characteristic frequencies and the classification of the SWs spectra for meander-film were described in Ref. [8]. Fig. 2 (b) shows the results of calculating the spectra of SW excitations magnetized along the $z$-axis at $\varphi=90^O$. The characteristic frequencies of the spectra for all values of the external magnetic field $H_{ext}$ are localized above the frequency of the uniform ferromagnetic Kittel resonance mode $f_\perp(H_{ext}) = \gamma\sqrt{\mu_0 H_{ext}(\mu_0 H_{ext} + M_s)}$ of the in-plane magnetized CoFeB film, [18]which is shown by the black dashed curve in Fig. 2 (b). Overall, a satisfactory agreement is observed between the experimental results of the VNA-FMR and the micromagnetic simulations.

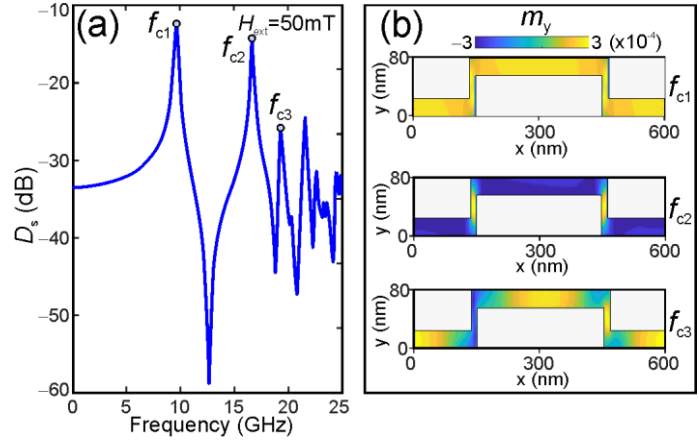

**Fig.3.** (a) Measured SW excitation spectrum at the value of the bias magnetic field $\mu_0 H_{ext}$=50 mT directed along the cores ($\varphi=90^O$). (b) Color-coded distribution of the dynamic magnetization components $m_y(x,y)$ within the unit cell of the meander-shaped CoFeB film for the three lowest frequencies modes $f_1, f_2,$ and $f_3$. Each panel corresponds to the value of the frequency indicated in panel (a). The frequencies $f_{c1}$=9.60 GHz, $f_{c2}$=16.56 GHz, and $f_{c3}$=19.24 GHz are indicated with closed circles.

Fig. 3 (a) shows the spectrum of characteristic frequencies calculated for a fixed external magnetic field $H_{ext}$ = 50 mT in the $x$-direction ($\varphi=0^O$). To understand the origin of the resonant modes observed in Fig. 3 (a), we extracted the distribution of the dynamic magnetization at the corresponding frequency of the observed modes. Fig. 3 (b) shows the spatial distribution of the dynamic magnetization ($m_y$) for the three strongest spectral peaks ($f_1, f_2$ and $f_3$). It can be seen that the lowest frequency mode ($f_{c1}$=9.60 GHz) corresponds to a quasi-homogeneous oscillation mode of dynamic magnetization throughout the entire meander-shaped film. The frequency $f_{c2}$=16.56 GHz corresponds to in-phase quasi-homogeneous magnetization oscillations in the horizontal segments with nodal points at the 90° corners between the vertical and horizontal segments. Finally, the frequency $f_{c3}$ =19.24 GHz corresponds to dynamic magnetization mainly concentrated in the horizontal segments but oscillating out-of-phase in the vertical segments.

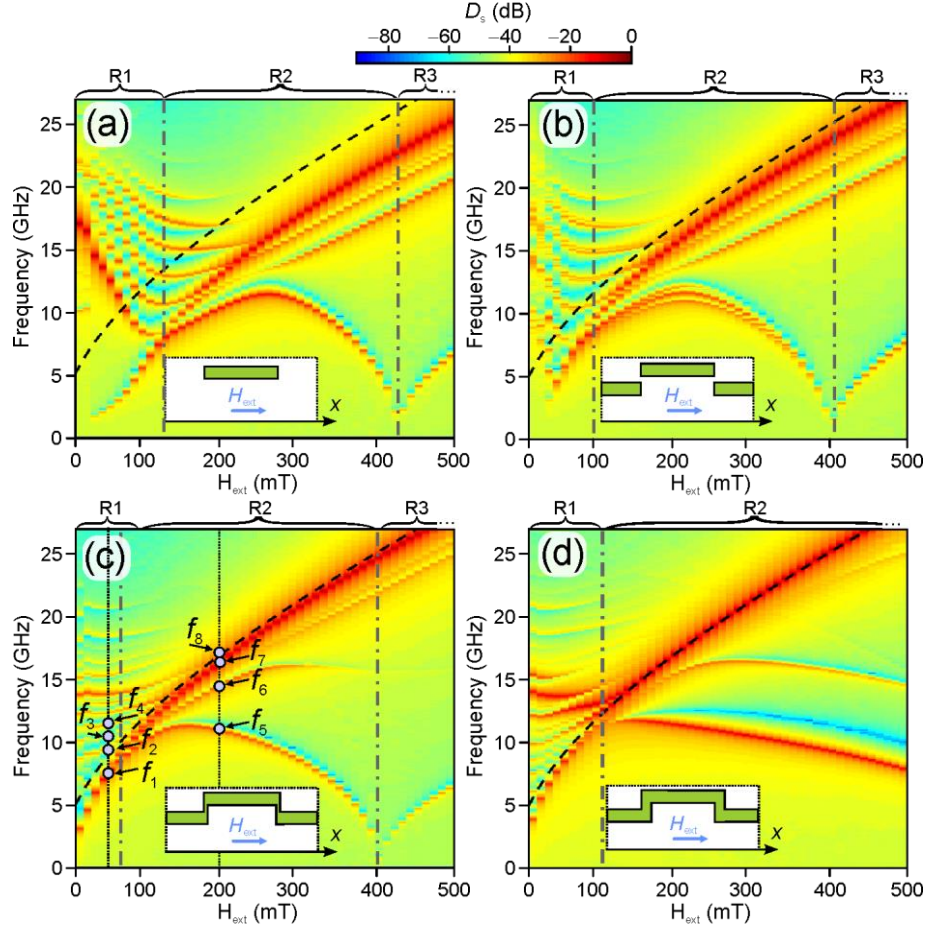

**Fig. 4.** Micromagnetic simulation results for the magnetic field dependence of the power spectral density for a periodic sequence of isolated stripes (a), the multilevel stripe array (b), the meander-shaped film at $d_2$=12 nm (c) (i.e the sample experimentally studied in the present work) and $d_2$=23 nm (d). All data are plotted for $d_1$=23 nm and the external magnetic field directed perpendicular to the cores ($\varphi$=0°). The sketch of the structures is shown in the insets. The frequencies for the spatial dynamic magnetization analysis are marked by circles: $f_1$=8.19 GHz, $f_2$=9.57 GHz, $f_3$=11.47 GHz and $f_4$=13.3 GHz at $H_{ext}$=50 mT and $f_5$=11.08 GHz, $f_6$=14.63 GHz, $f_7$=16.97 GHz and $f_8$= 17.39 GHz at $H_{ext}$=200 mT. The dashed line marks the FMR frequency $f_\perp(H_{ext})$ corresponding to the uniform precession in planar 23 nm thick CoFeB, the main mode or the corresponding points in the experimental data.

To fully understand the field-dependent frequency and the type of the SW excitation observed in the meander-shaped film, we plot in Fig. 4 the calculated SW spectra for different types of structures, from a single isolated magnetic stripe to more complex structures of the vertical meander-shaped film. This helps us to understand the transformation of the SW spectra starting from a separate array of CoFeB stripes (Fig.4 (a)) and a multilevel stripe array (Fig. 4 (b)) to the meander-shaped structure (Fig.4 c, d) magnetized along the *x*-axis. The vertical gap (edge-to-edge distance) was 30 nm for the multilevel stripe arrangement, *i.e.* it corresponds to the experimental design of the meander structure, but without vertical segments ($d_2$=0). The field-frequency map ($H_{ext}$, $f$) can be divided into three field regions, labeled R1, R2 and R3 at the top of each panel in Fig. 4. The first region R1 is related to the non-uniform magnetic field profile within each stripe which manifests itself in the domains oriented antiparallel along *y*-axis and localized close to the stripe edges limiting each stripe in the *x*-direction.[19] Within region R1 the average magnetization inside the stripe is oriented along z-axis (see Fig.2a from Ref. 13). Second region R2 for the stripes does still consist of these localized modes in the edge domains. At the same time, the central modes with the quasi-uniform precession and dispersion similar to $f_{perp}$ ($H_{ext}$) would correspond to region R2. In addition to the central mode (of quasi-uniform precession amplitude over the stripe width) localized modes, whose frequency decreases with the increase of the external field. At a critical field $H_{crit}$, defined by the condition $B_z$=0 and $M_z$=0 in the whole volume of the sample, this low-frequency mode disappears and, in the region R3 for H>$H_{crit}$, its behavior is similar to that of the main mode, the precession being quasi uniform distributed over the unit cell. This low-frequency mode behaves like a resonant backward volume spin wave [20].

To understand the impact of the vertical segments on the magnetization dynamics and mode profile we have simulated additional structures with increasing degree of complexity, i. e. from isolated to multilevel stripes until meander-shaped shaped films with different widths of the vertical segments.

The multilevel stripe array (Fig.4b) can be considered as the intermediate step between isolated stripes (Fig. 4 (a)) and meander structure (Fig. 4 ((c) and (d)). It can be noted, that in this case the parabolic dependencies of modes in the region R1 are is still present but the frequency minima move toward lower fields than the case of Fig. 4 (a). At the same time, this is quite evident that the main mode dispersion is closer to the FMR frequency $f_{\perp}\left(H_{ext}\right)$ inside the region R2 and R3 since the mean value of the magnetization is more similar to the uniform CoFeB film. This situation is different from the stripe array for the dynamic magnetization as well since multilevel array can exhibit the coupling between spin waves oscillation inside each stripe. This collective SW dynamics manifest itself in the multiple branches of the edge modes in R2 with frequencies lower than that of the quasi-uniform.

The meander-shaped film behavior is very similar to the SW excitation spectra observed for the stripes without vertical sections (Fig.4c). In this case, the edge mode branches exhibit an almost monotonically increasing frequency with the increase of the field in R1, which can be compared with the pronounced minima in Fig. 4 (c) and Fig. 4 (d). It should be noted that this mode is not visible in the experiment shown in Fig. 2 (c), as the dispersion of this mode is strongly affected by a small misalignment of magnetization with respect to the groove length. This is demonstrated in Fig.5 where we show that a small angular variation of φ from $0^O$ to $5^O$ leads to a smoothing of the boundary between the R2 and R3 regions due to misalignment of the static magnetization to the *z*-axis and the flattening of the low-frequency spin-wave branch. Moreover, the increase in the thickness $d_2$ of the vertical sections of the meander leads to an increase in $H_{\text{crit}}$. This can be attributed to the increase of $H_{\text{crit}}$ where the average magnetization in the z-direction decreases to the value $M_z=0$ with the increase in the volume of the vertical section. The results presented in Fig. 5 are important since they show how critical is the slightly misaligned of the applied field with respect to the groove's periodicity direction. (φ=$0^O$) for the experimental determination of the minimum frequency for the mode with the lowest frequency

The SW excitation spectra for isolated vertical segments of the meander structure are shown in Fig.S1 of the Supplementar material, where also the additional data related to isolated segments are collected.[20]

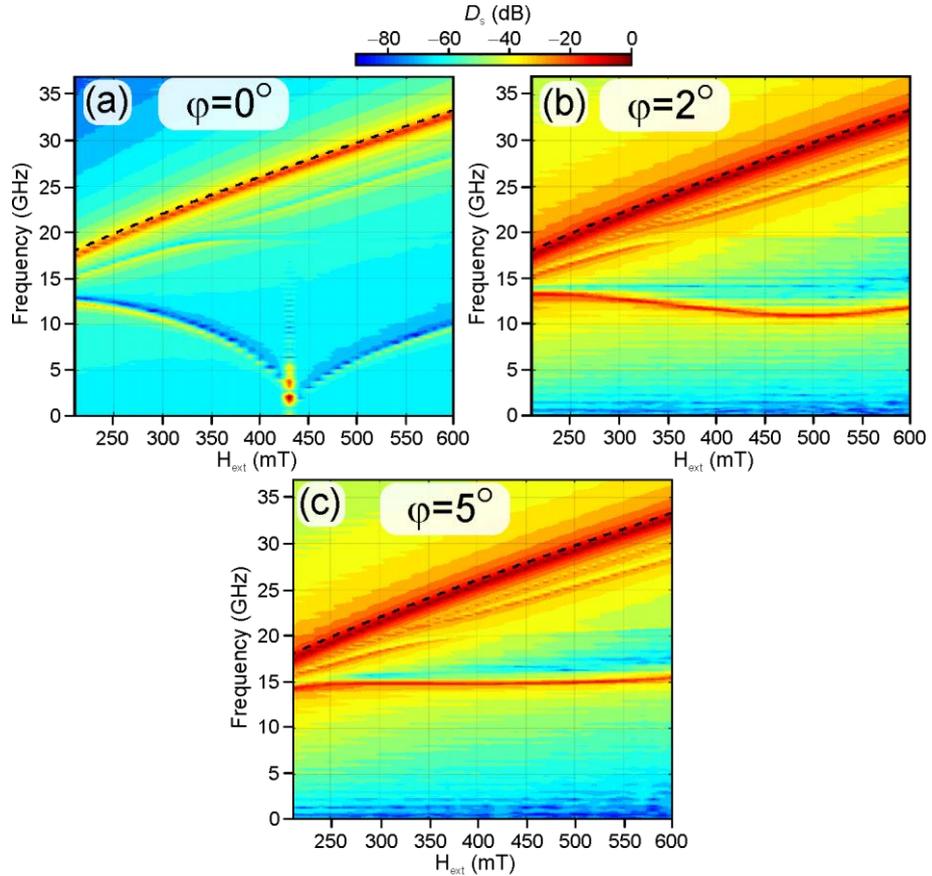

**Fig. 5.** Micromagnetic simulation of the spectra of SW excitations in meander-shaped structure at φ=$0^O$, φ=$2^O$ and φ=$5^O$ demonstrating the high sensitivity of the frequency minimum for the lowest frequency mode on the small angular variation .

Next, after consideration of the evolution of spectra based on the variation in stripes' geometry, we perform the simulation of static magnetization profile along the ξ coordinate inside the meander structure (see Fig.6). From the analysis of the evolution of static magnetization components profile along with the curvilinear coordinate it is seen that with the increase of $H_{ext}$ only the $M_z$ component significantly transforms. The x-component of magnetization is non-zero in horizontal sections whilst y-component has the positive and negative sign in vertical segments in the whole range of $H_{ext}$ extending from 50 mT up to 500 mT. Here we can emphasize that the main difference between the localized modes in the region R1 shown in panels of Fig.4 for stripes and meander shaped structure is associated with the localized oscillation of magnetization at the edges of the stripe (Fig.4 a and b) and within the vertical section of meander structure (Fig.4 c,d). At the same time, the highly nonuniform $M_z$-profile below the critical field $H_{crit}$ in the vicinity of the vertical section is responsible for the mode dispersion between the quasi-uniform mode and low frequency mode inside the region R2. The frequency of these SW excitation branches increases for stripes (Fig.4 a, b) and decreases for the meander-shaped stripes (Fig.4c,d). The static magnetization for the stripe is strongly nonuniform in z-direction while for meander the dominant orientation in the horizontal sections is along x-axis. If we consider the central stripe of multilevel stripe array the orientation of static magnetization tends to be aligned along x-direction as in meander-shaped film but still is highly nonuniform in z-direction at values of the magnetic field below 500 mT.

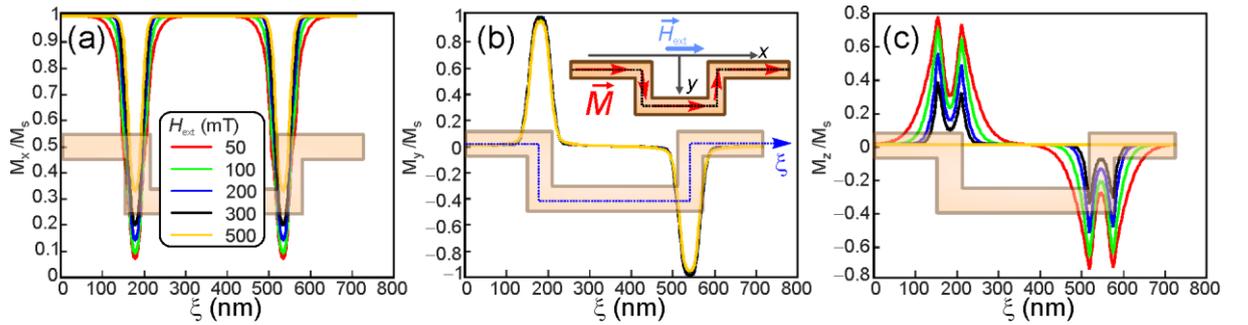

**Fig. 6.** The evolution of static magnetization components profile (a) – $M_x$, (b) - $M_y$ and (c) - $M_z$ along the curvilinear coordinate ξ enveloping the meander structure with the increase of the value of the external magnetic field $H_{ext}$ from 50 mT to 500 mT. The schematic of meander the structure is shown at each panel demonstrating the vertical orientation of static magnetization inside the vertical segments.

To demonstrate how the antiparallel orientation of the magnetization inside the vertical section manifests itself in the spin-wave excitation spectra we perform the micromagnetic simulation of the dynamic magnetization profile for applied magnetic fields within the R1 and R2 regions. Figs. 7 and 8 demonstrate the spin-wave FMR spectra and spatial distributions of the $m_x$, $m_y$ and $m_z$ component of magnetization over the cross-section of the unit cell at different frequencies (frequency values are shown at the right parts of each panel).

Fig. 7(a) shows the spectra of SW excitations for the meander-shaped structure magnetized across cores (φ=0°) at $H_{ext}$=50 mT. For external magnetic fields lower than 150 mT (see Fig. 6), the equilibrium state of magnetization in the meander is substantially inhomogeneous. The spectrum at $H_{ext}$<150 mT contains spectral peaks at frequencies above the frequency $f_\perp(H_{ext})$. Fig. 7(b) demonstrates the dynamic magnetization distribution for the first four modes presented in the spectra. It should be noted that $m_x$-component is an order of magnitude smaller than $m_y$ and $m_z$, and the oscillation of $m_x$ occurs also in horizontal sections but in the region close to the vicinity of the vertical section for all four modes. The number of nodes in the $m_y$ and $m_z$ spatial profiles increases with the frequency increase from $f_1$ up to $f_4$.

For fields $H_{ext}$> 150 mT, the magnetization distribution is close to homogeneous, and below the frequency $f_\perp(H_{ext})$, there are modes corresponding to magnetization oscillations generated by backward volume waves in horizontal segments, localized modes and modes formed by forward volume waves in vertical segments [20]. From the spin-wave excitation spectra for $H_{ext}$=200 mT (see Fig.8a) one can note, that the peak at frequencies $f_7$ and $f_8$ are broadened due to high order modes contribution. This fact can be clearly demonstrated from the mode profile at the frequency $f_8$ (see Fig.8b), where seven nodes in $m_y$-distribution are distinguished. Thus, the profile at $f_7$ has the modulation in the horizontal section of the meander. This modulation does not have the form of a standing wave since the absence of alternating positive and negative values in $m_y$-profile. This non-uniform profiles of quasi-uniform mode at $f_7$ can be attributed with the hybridization of the quasi-uniform mode with the almost degenerated spectra of the localized modes, well established in the region R1. In the range R2 the $m_x$-component are localized only in vertical sections and oscillates in anti-phase in both vertical segments. Such in-phase and out-of-phase precessional mode of dynamic magnetization have already been observed in the past in both isolated [21] and dense [22] NiFe/Cu/NiFe planar stripe arrays because of the dipolar coupling between the two ferromagnetic layers.

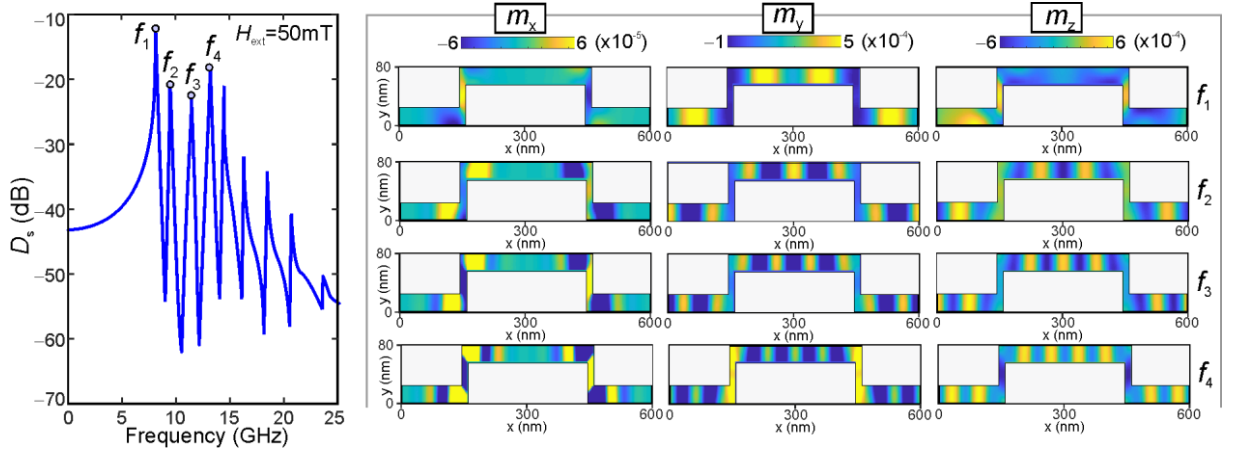

**Fig. 7.** (a) SW excitation spectra at the value of bias field $\mu_0 H_{ext}$=50 mT applied perpendicular to the cores ($\varphi=0^O$). (b) Color-coded distribution of the dynamic magnetization components $m_{x,y,z}$ $(x,y)$ inside the unit cell of magnonic meander film at the peak frequencies from $f_1$ to $f_4$. Each row corresponds to the value of frequency denoted at the right part of the panels.

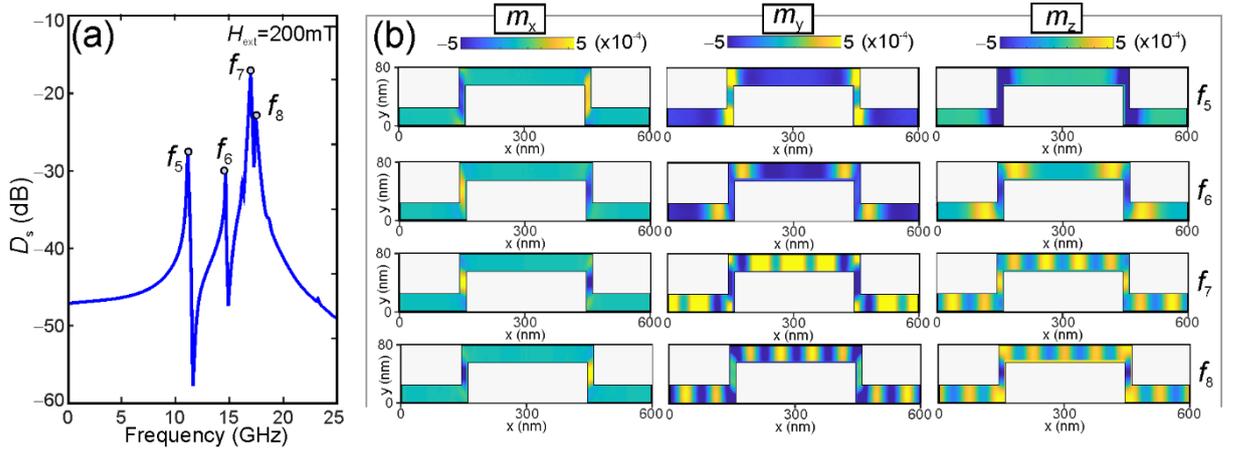

**Fig. 8.** (a) Spin-wave excitation spectra at the value of bias field $\mu_0 H_{ext}$ =200 mT applied perpendicular to the cores ($\varphi=0^O$). (b) Color-coded distribution of the dynamic magnetization components $m_{x,y,z}$ $(x,y)$ inside the unit cell of magnonic meandering film at the peak frequencies from $f_1$ to $f_8$. Each row corresponds to the value of frequency denoted at the right part of the panels.

The modes at $f_5$ and $f_6$ have an in-phase oscillation of $m_y$ and $m_z$ components in the vertical segments of the meander structure and strong localization in the vertical segments and in the part of the horizontal section close to the transition region being interconnection between vertical and horizontal sections. As it was emphasized before, the associated frequencies for these modes decrease with the increase of $H_{ext}$. For mode $f_6$ this is peculiar difference with the behavior of such mode in the case of stripe/multilevel array structure.

**Conclusion**

In conclusion, the spin-wave excitation spectra for meander-shaped CoFeB thin films were studied using the broadband FMR-VNA technique and micromagnetic simulations. We measure the spin-excitation frequency dependence on the applied field strength applied along the easy and hard magnetization direction of the meandering film. Using micromagnetic simulations we reproduce the experimental findings and derive information on the spatial character of spin-wave modes formation inside the meander structure. It was shown that the meander-shaped structure has advantages over the 1D magnetic stripe array exhibiting rich spin-wave mode spectra along with the in-phase and out-of-phase oscillation regimes inside the vertical parts of the meander. Tunability of the resonance characteristics can be performed by the means of $H_{ext}$ value and bias angle. The latter could be potentially used in the magnetic memory based on the profiled magnetic films and in the fabrication of tunable magnetic metasurfaces.

**Acknowledgments**

Numerical micromagnetic simulation of spin-wave excitation in meander-shaped film and comparison with experimental data was supported by Russian Science Foundation (project #20-79-10191). S.A.N. acknowledge of Russian Science Foundation (project #19-19-00607), which supports the calculation of static magnetization profile in magnetic structure. Imec's contribution to this work has been supported by its industrial affiliate program on beyond-CMOS logic as well as by the European Union's Horizon 2020 research and innovation program within the FET-OPEN project CHIRON under Grant Agreement No. 801055.

* A.V. S and G.T contributed equally to this work.